\documentclass[12pt,final]{iopart}
\usepackage{iopams}  

\usepackage{xcolor}
\usepackage{graphicx}

\expandafter\let\csname equation*\endcsname\relax

\expandafter\let\csname endequation*\endcsname\relax

\usepackage{amsmath}
\usepackage{mathrsfs}
\usepackage[T1]{fontenc}
\newcommand{\thorn}{\mbox{\th}}
\usepackage{float}
\usepackage{tikz}
\usetikzlibrary{decorations.pathmorphing}
\usetikzlibrary{arrows.meta}
\usetikzlibrary{external}

\usepackage{xspace}
\usepackage[textsize=footnotesize]{todonotes}

\usepackage{subcaption}

\usepackage{showkeys}
\usepackage{hyperref}

\usepackage[backend=biber,
            style=phys,
            sorting=nyt,
            articletitle=true,
            biblabel=brackets,
            chaptertitle=true,
            pageranges=true,
            hyperref=true,
            maxcitenames=3,
            maxbibnames=4,
            url=true,
            autocite=superscript
            ] {biblatex}

\newcommand{\eps}{\varepsilon}

\newcommand*{\hdal}{{\Box\kern-7.5pt\wedge}}

\newcommand*{\scri}{\mathscr{I}}

\newcommand*{\ii}{\mathrm{i}}

\renewcommand{\Im}{\mathrm{Im}}

\begin{document}
\title[The non-linear perturbation of a black hole by gravitational waves]{The non-linear perturbation of a black hole by gravitational waves. II. Quasinormal modes and the compactification problem}

\author{J Frauendiener$^1$ and C Stevens$^2$}

\address{${}^1$Department of Mathematics and Statistics, University of Otago, Dunedin 9016, New Zealand}
\address{${}^2$School of Mathematics and Statistics, University of Canterbury, Christchurch 8041, New Zealand}

\ead{joergf@maths.otago.ac.nz, chris.stevens@canterbury.ac.nz}

\begin{abstract}

Recently, Friedrich's Generalized Conformal Field Equations (GCFE) have been implemented numerically and global quantities such as the Bondi energy and the Bondi-Sachs mass loss have been successfully calculated directly on null-infinity. Although being an attractive option for studying global quantities by way of local differential geometrical methods, how viable are the GCFE for study of quantities arising in the physical space-time? In particular, how long can the evolution track phenomena that need a constant proper physical timestep to be accurately resolved? We address this question by studying the curvature oscillations induced on the Schwarzschild space-time by a non-linear gravitational perturbation. For small enough amplitudes, these are the well approximated by the linear quasinormal modes, where each mode rings at a frequency determined solely by the Schwarzschild mass. We find that the GCFE can indeed resolve these oscillations, which quickly approach the linear regime, but only for a short time before the compactification becomes ``too fast'' to handle numerically.
\end{abstract}
% \keywords{numerical relavitiy, conformal field equations, quasinormal modes}
\submitto{\CQG}
\maketitle

\section{Introduction}
Recently, Friedrich's Generalized Conformal Field Equations (GCFE) (see \cite{Frauendiener:2004a} and \cite{ValienteKroon:2016} for comprehensive reviews of the conformal field equations and their applications), which are a regular extension of Einstein's field equations to the conformal boundary, have been implemented numerically as an Initial Boundary Value Problem (IBVP)~\cite{Beyer:2017}. As it is the case with the standard Einstein equations, the GCFE also have to be augmented with gauge conditions. While in the standard case, this amounts to imposing conditions for the coordinates and, possibly, a tetrad, in the case of the GCFE there is an additional choice of a conformal factor and a conformal connection. The implementation in~\cite{Beyer:2017} made use of the simplest and most natural way to fix these choices, namely the conformal Gauß gauge (CGG) which makes use of time-like conformal geodesics. It is similar in nature to the well-known Gauß gauge but has the advantage that it covers the global Schwarzschild and Kerr space-times. This framework has been used successfully to calculate global quantities such as the Bondi energy directly on null-infinity ($\mathscr{I}^+$) \cite{Frauendiener:2022,Frauendiener:2021a}, produced by the fully non-linear response of a Schwarzschild black hole to an ingoing gravitational wave.

While the main purpose of the numerically implementation of the GCFE with the conformal Gauß gauge in~\cite{Beyer:2017} is to calculate global quantities directly on the conformal boundary, it is of interest to see how well they can perform in the physical region as well. In particular, issues were brought to attention in~\cite{Frauendiener:2021a} regarding how the compactification procedure hinders evolutions with a constant physical timestep for an arbitrary length of physical time. How suited then is this implementaion to study local-in-physical-time phenomena?

To investigate this, we further investigate the prominent curvature oscillations that were observed in~\cite{Frauendiener:2021a}, of which the linear equivalents are the Quasinormal Modes (QNM). These have been thoroughly studied in the literature (see \cite{Kokkotas:1999} and references within) and will be our test case. They have a frequency dependent only upon the mass, angular momentum and charge of the black hole, and in particular, are independent of the perturbation. The Teukolsky equation \cite{Teukolsky:1973,Press:1973,Teukolsky:1974} is the master equation that describes the scalar, electromagnetic or gravitational perturbations of the Kerr space-time. The quantities that represent gravitational radiation are given by $\Psi_0$ (spin $2$) and $\Psi_4$ (spin $-2$) with respect to a suitably chosen null-tetrad, with the equations governing their perturbations decoupled from each other and the other components and involving only quantities from the background space-time. We are interested in the Schwarzschild space-time, which is easily obtained from the above by setting the angular momentum to zero.

In linear theory around Schwarzschild space-time, the frequencies of the QNM are computed for static observers at a fixed radius with respect to the Schwarzschild time coordinate. As we are in the non-linear regime, we need to have a reasonable way to compare the two cases. Since the radial Schwarzschild coordinate is defined by the radius of the spheres of the spherical symmetry, we select curves, world-lines of ``quasi-static observers'',  which lie on coordinate spheres of constant area at fixed angles (see below). Along both the Schwarzschild curves and our ``quasi-static observers'' we use the proper time to compute the frequencies. It is along these curves that we compare the observed curvature oscillations to QNM. 

In linear theory, it is standard procedure to assume the perturbation to be supported outside the horizon. In contrast, there is nothing to stop us from choosing quasi-static world-lines inside the black hole. Furthermore, we can identify generators along null-infinity and compute the so-called \emph{Bondi time} along them, which corresponds to the physical proper time of an observer at infinity~\cite{Frauendiener:2000}. Thus, the non-linear curvature oscillations can be observed along two different types of curves that lie outside the standard approach of linear theory.

Using the fully non-linear curvature oscillations, we see how far in physical proper time they can be accurately resolved before compactification makes this impractical numerically. To postpone compactification for an arbitrary length of time would be very attractive as it would allow one to study both local and global behaviour in the same framework. Potential avenues to accomplish this are put forward and discussed.

The paper is organized as follows: in Sec.~\ref{sec:system} we give a brief overview of the implementation of the GCFE as a numerical initial boundary value problem. Defining the quasi-static world-lines and the Bondi time in Sec.~\ref{sec:DetailedComparisons}, we also compare the frequency of the curvature oscillations to the Quasinormal Frequencies (QNF) of linear theory along a variety of curves for different ingoing wave pulses. In Sec.~\ref{sec:ProblemsandResolutions} we discuss the different ways one may attempt to extend the amount of physical proper time that can be resolved. Sec.~\ref{sec:summary} contains a summary of the results and discusses future work. 

% In all of these cases, the gravitational perturbation of the specific mode induces perturbations in the other modes, which we also use for comparison. We find strong agreement even in the non-linear regime to the linear theory at lower frequencies, while the higher frequencies had a larger deviation. Secondly, we present results of an in-depth investigation into the issues faced when simulating phenomena where a constant physical timestep is required.

\section{The system}\label{sec:system}
Here we give a brief overview of the GCFE system. We refer the reader to~\cite{Beyer:2017,Frauendiener:2021a} for a detailed exposition.

\subsection{The conformal field equations}

We use the general conformal field equations as defined by Friedrich~\cite{Friedrich:1995}. This is a system of PDEs which characterizes a vacuum space-time in terms of its conformal structure. Thus, it implements Penrose's proposal~\cite{Penrose:1965} for the conformal compactification of space-times and it gives direct access to the conformal boundary, null-infinity, of the space-time. In order to do evolutions with these equations one needs to set up an IBVP and this involves the choice of gauge conditions. In our implementation we opted for the simplest choice, namely the so called conformal Gauß gauge as introduced by Friedrich~\cite{Friedrich:2002}. This gauge is a generalization of the standard Gauß gauge, but it is adapted to time-like conformal geodesics \cite{Tod:2012} rather than time-like metric geodesics. It is well-known that the standard Gauß gauge is a terrible choice for numerical evolutions because the geodesics tend to converge and, ultimately, intersect thereby destroying the regularity of the attached coordinate system. In contrast, the conformal geodesics are governed by equations containing an acceleration term which, in turn, is determined by the curvature. Friedrich has shown that this gauge yields a regular semi-global covering of the Schwarzschild space-time~\cite{Friedrich:2003}, thus allowing access to null-infinity. Since we study space-times close to Schwarzschild we expect that the gauge maintains these properties as long as we stay ``close enough''.

We write the GCFE in the space-spinor formalism (see~\cite{Sommers:1980,Beyer:2017,Frauendiener:2021a}), mainly because the irreducible decompositions of the involved spinor fields yield concise subsets of equations and because it is very closely related to splitting the equations into two systems of evolution and constraint equations. Furthermore, we foliate the hypersurfaces of constant time into by nested 2-surfaces of spherical topology. This allows us to use the $\eth$-formalism of Newman and Penrose~\cite{Newman:1966} together with a pseudo-spectral method based on the so-called spin-weighted spherical harmonics ${}_sY_{lm}$ (SWSH) as described in~\cite{Penrose:1984a}. These choices yield a first-order evolution system which has been shown to propagate the constraints analytically and in a numerically well-behaved manner. In addition, due to the conformal Gauß gauge, most system variables propagate along the time coordinate lines, the exception being the rescaled Weyl spinor, for which the equations become a symmetric hyperbolic, i.e., wave-like, system of PDEs. This system is referred to as the Bianchi equation because it finds its origin in the Bianchi identities for the curvature when the Einstein equation is imposed. This agrees with the fact that this quantity describes the gravitational wave degrees of freedom in the space-time.

We solve the system as an IBVP with time-like boundaries (see Fig.~\ref{fig:contourplot} as an example of the IBVP applied to the Schwarzschild space-time), where constraint preserving boundary conditions, as presented in~\cite{Beyer:2017}, are used to prevent constraint violating modes from entering the domain. An analysis of the system yields five characteristic variables, of which two are ingoing, two are outgoing, and one propagates along the boundary. At a given boundary, of the two ingoing modes, one propagates along light cones while the other is time-like. We prescribe arbitrary boundary data for the light-like variable, which represents the one complex physical degree of freedom in the Einstein equations, namely the ingoing gravitational radiation, while the other is fixed uniquely by the requirement that no constraint violating modes enter into the computational domain. We should point out here that the relative simplicity of formulating the constraint preserving conditions is mainly due to our use of the conformal Gauß gauge which confines the wave-like behaviour to the only physically relevant fields. This is the unique advantage of this choice. Any other gauge choice would yield a much more complicated boundary procedure.

\subsection{Perturbations of the Schwarzschild space-time}

\begin{figure}[H]
    \centering
    \includegraphics[width=0.6\linewidth]{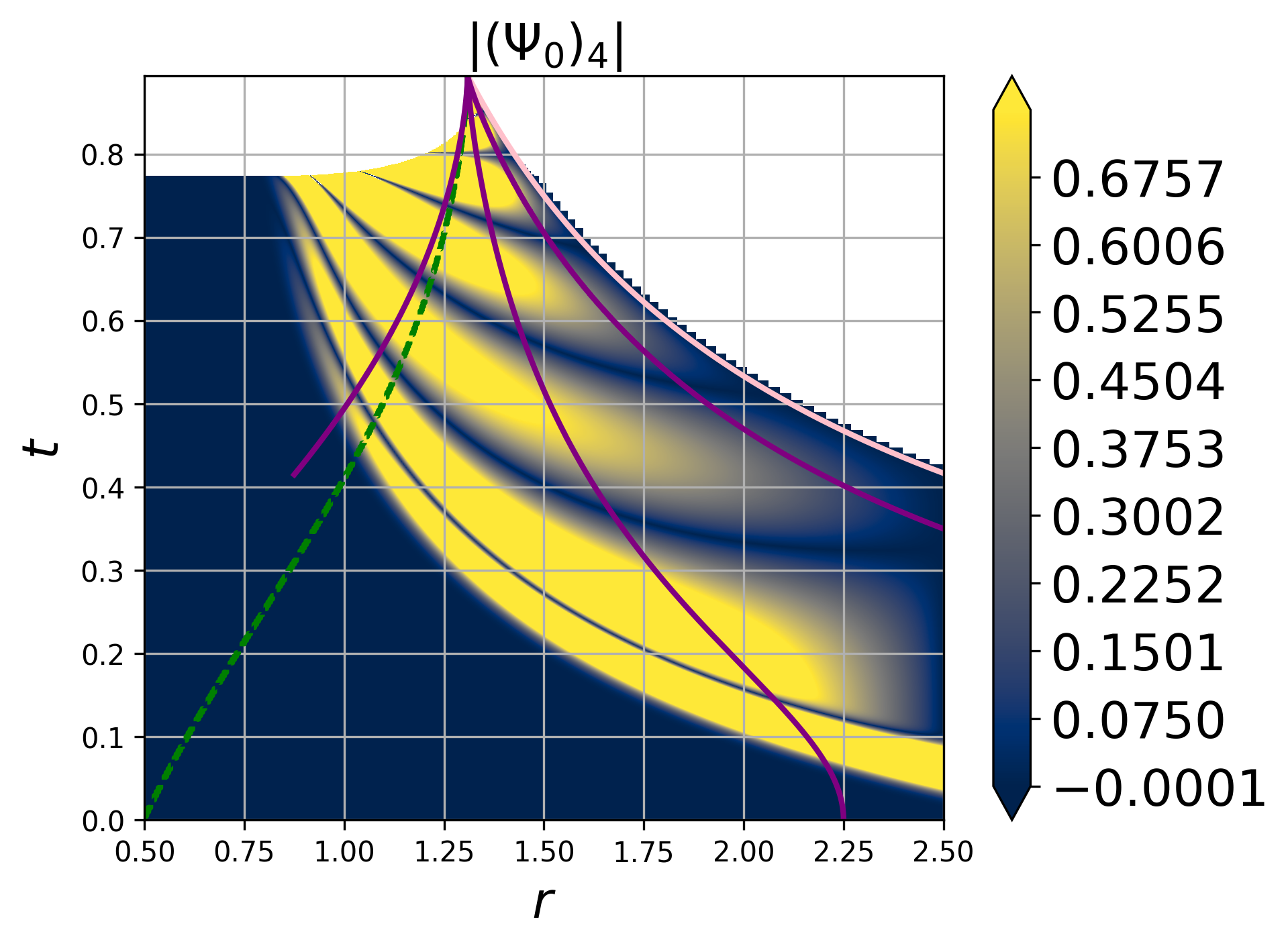}
    \caption{A contour plot of the magnitude of the $l=4$ spectral coefficient of $\Psi_0$. The green curve is an approximation to the apparent horizon, the pink curve is null infinity and the purple curves are \emph{quasi-static world-lines}. One can clearly see the ingoing wave $\Psi_0$ travelling in from the right boundary, after which the curvature oscillations begin.}
    \label{fig:contourplot}
\end{figure}

We study the non-linear perturbation of the Schwarzschild space-time by gravitational waves. Within our frame-work this can be accomplished in a very clean way by prescribing exact Schwarzschild initial data. We use isotropic coordinates $\{t,r,\theta,\phi\}$, and use a conformal factor which compactifies the spatial initial data surface as given in~\cite{Friedrich:2003}. The perturbation of the space-time is created by an ingoing pulse of gravitational radiation from the outer boundary. In axi-symmetry this pulse is a linear combination of SWSH basis functions ${}_2Y_{l0}$ for various choices of $l$. The pulse propagates into the domain and passes through the horizon into the black hole. After a short time, the black hole responds by oscillations which can be seen in the system variables. Null-infinity is reached and the system evolves until it is very close (with respect to conformal time) to time-like infinity, where the fields diverge. See Fig.~\ref{fig:contourplot} for a contour plot of the magnitude of the $l=4,\;m=0$ spectral coefficient of $\Psi_0$ that showcases the situation.

\subsection{Numerical setup}

We use the MPI-parallelized Python package COFFEE \cite{Doulis:2019} to evolve the system. It contains a large selection of time integrators, finite difference operators with the summation-by-parts property as well as the ability to use the pseudo-spectral code presented in \cite{Beyer:2015,Beyer:2016}. This allows us to perform semi-analytic angular derivatives through a fast implementation of the $\eth$-operators optimized for axi-symmetry. We use the fourth order explicit Runga-Kutta method for time integration and Strand's summation-by-parts finite difference operator given in \cite{Strand:1994} for radial derivatives, which is fourth order accurate in the interior and third order accurate close to the boundaries. Finally, the simultaneous approximation term (SAT) method as presented in~\cite{Carpenter:1999} is used to impose stable boundary conditions. While the code can operate in full $3+1$ mode, for the present discussions we assume axi-symmetry so that the only non-vanishing SWSH are the ones with $m=0$. This greatly reduces the computational expense while still permitting non-trivial deformations of the black hole space-time.

We discretize the radial grid into $401$ equidistant points in the interval $[m/2,\,m/2 + 2m]$, with $m$ the Schwarzschild mass of the initial space-time, so that initially the inner boundary is on the event horizon (recall that $r$ is the isotropic radial coordinate). The angular grid is discretized into $33$ equidistant points in the interval $[0,\pi)$ and we truncate the SWSH expansion at $l=21$. This allows for the spectral coefficients of all system variables to reach machine precision even at late times.

%We assume, as usual, that space-time is foliated into 3-dimensional time-slices which, in turn, are foliated by topological 2-spheres. This allows us to take advantage of pseudo-spectral methods using Spin-Weighted Spherical Harmonics (SWSH)~\cite{Penrose:1984a}. 

We use an adaptive timestep given by $\Delta t = (CFL/C(t)) \Delta r$, where $CFL$ is the Courant-Friedrichs-Lewy number that we take to be $0.2$, and $C(t)$ is the maximum value of all the characteristic speeds evaluated on the entire spatial grid at a given time. This ensures that we do not allow the system to propagate over multiple radial grid points in one timestep, thus avoiding the associated instabilities.

We compute the radius $r_{OTS}$ of the largest grid sphere which is completely trapped (see~\cite{Frauendiener:2021a} for details) to make sure that the inner boundary of the computational domain is always inside the trapped region so that no ingoing boundary conditions need to be specified on that boundary. This will deviate from $r=m/2$ once perturbed by the ingoing gravitational wave. As this grid sphere grows into the interior of the computational domain during the evolution (see Fig.~\ref{fig:contourplot}), we perform a \emph{regridding}, by cutting out a chunk of the space-time inside this sphere. One condition for regridding could be to stop the inner boundary from going no further than a certain number of grid spheres from $r_{OTS}$. However, as we are interested in looking at quasi-static observers inside the black hole, we want to include a sufficient amount of space-time here. So instead, the condition to regrid the inner boundary is given by when the components of the rescaled Weyl spinor $\psi_{ABCD}$ diverge beyond a specified value, which we take to be $10^5$. This then avoids the singularity, and proves to leave sufficient space-time inside the black hole for our purposes. Interpolation of the system variables and gauge quantities is performed to obtain values on the radially smaller, but still equidistant spatial grid with the same number of grid points as before. This procedure is performed analogously in the neighbourhood of the outer boundary once $\mathscr{I}^+$ has moved sufficiently into the computational domain.

We impose exact Schwarzschild boundary conditions on the inner boundary for the entire simulation which yields the best stability properties for the numerics and still allows satisfaction of the constraints inside the black hole. On the outer boundary, we impose our wave profile (defined in Sec.~\ref{sec:DetailedComparisons} by Eq.~\ref{eq:WaveProfile}) until we reach $\mathscr{I}^+$. From that point onwards, the goal on this boundary is to continue to evolve in a stable fashion as close to time-like infinity $i^+$ as possible. Only numerical errors can propagate from the boundary onto $\mathscr{I}^+$ or the physical domain, and these can be minimized with increasing resolution. Thus, we do not impose any boundary conditions at this point but instead enforce the evolution equations. This has proved to be the most stable approach to enable synergy with the regridding procedure.

Convergence tests for the above numerical procedures have been done in \cite{Beyer:2017,Frauendiener:2021a} and we just mention here that all constraints converge at the correct order everywhere, including on $\mathscr{I}^+$ and inside the black hole, even when we are within a very small interval of conformal time of $i^+$.

\section{Comparison of linear and non-linear ringing}
This section details how we compare the QNF of linear perturbation theory for the Schwarzschild space-time to our non-linear results. Since in our simulations we are dealing with a non-linear evolution, there is no fixed static and spherically symmetric background space-time which could be used as a reference. Thus, we need to select curves in our numerically generated space-time that mimic the curves given by constant Schwarzschild radii. These curves are parameterized by the physical proper time, and we estimate the frequency of the oscillating components of the (physical) Weyl curvature $\Psi_{ABCD}$. We present comparisons for a variety of different SWSH modes of the $\Psi_i$ for the two cases where the ingoing wave is proportional to ${}_2Y_{20}$ or ${}_2Y_{40}$.

\subsection{Meaningful comparisons between linear and non-linear regimes}

The QNFs of the linear theory are given with respect to the Schwarzschild time coordinate which is the proper time of the observer at infinity. In order to compare with the non-linear case, we transform these frequencies to frequencies with respect to the proper time of the static observers at constant Schwarzschild radius. Schwarzschild time can easily be transformed to physical proper time $\tau$ by the scaling $\tau = \sqrt{1 - 2m/\tilde{r}}\,\tilde{t}$, where $\tilde{r}$ is the Schwarzschild radius.

In the non-linear space-time, there are no static observers but we can introduce ``quasi-static observers'' which are defined as follows: on a given time-slice choose a point on an arbitrary grid-sphere and compute the area $A$ of that sphere. If the space-time was the exact Schwarzschild space-time, the Schwarzschild radius would then be $\sqrt{A/(4\pi)}$. One can then construct 2-surfaces on the subsequent time-slices by the condition that the area remains constant. Thus, we obtain a temporal sequence of 2-surfaces that have the same area. On each of these spheres we locate the point with the same angular coordinates as the starting point, obtaining a world-line that mimics a curve of constant Schwarzschild radius in the non-linearly perturbed space-time. The tangent vector to such a curve is given in the code coordinates by $\dot\gamma(t) = \{1,b(t),0,0\}$. We compute $b(t)$ by the requirement that the derivative of the area $A$ in the direction of the curve vanishes. Then, we can compute the physical proper time $\tau_1$ from a given timeslice, say $t=t_0$, to some future surface $t=t_1$ by
\begin{equation}
    \tau_1 = \int_{t_0}^{t_1} \sqrt{\tilde{g}_{00} + 2\tilde{g}_{01}b + \tilde{g}_{11}b^2}\,\text{d}t
\end{equation}

Comparison curves now identified, the final step is to estimate the frequency of the relevant quantities along them. This can easily be accomplished by identifying peaks and troughs of the oscillations and then using half-period lengths to approximate the frequency.

\subsection{Bondi time on null-infinity}

The quasi-static observers defined in the previous subsection pertain to time-like physical curves and the ability to compute proper physical time along them. For null generators of $\mathscr{I}^+$ we can use the notion of Bondi time~\cite{Penrose:1986}, which is the analogy to physical proper time on $\mathscr{I}^+$~\cite{Frauendiener:2000}. A Bondi time parameter $u$ is defined as a solution to the conformally invariant equation
\[
	\thorn_c'^2 u = 0,
\]
where $\thorn_c'$ is the conformally invariant $\thorn'$ operator of the GHP formalism \cite{Penrose:1986}. Expanding this equation yields
\[
	\Big{(}\thorn' - 2\rho'\Big{)}\thorn'u = \Big{(}D' - \eps' - \bar{\eps}' - 2\rho'\Big{)}D'u = 0,
\]
where $D' := N^a\nabla_a$, $N^a$ is tangential to the null generators and the spin-coefficients are computed in a Bondi frame as described in \cite{Frauendiener:2021a,Frauendiener:2022}. The solution $u$ is clearly of an exponential nature, and it will inherently have the property that $u\rightarrow\infty$ near $i^+$.

% Noting that in general $N^a$ will contain four non-zero components in the conformal space-time's holonomic basis $\{\partial_t,\partial_r,\partial_\theta,\partial_\phi\}$, we can take $t$ as the parameter along $\mathscr{I}^+$. As a consequence of this together with projecting these equations onto the submanifold that is $\mathscr{I}^+$, the $r$-derivatives are not required.
We regard $\mathscr{I}^+$ as parameterised by the conformal time $t$ and the angular coordinates. This allows us to write $N^a$ as a linear combination of $\partial_t$ and the angular derivative operators $\eth$ and $\eth'$ with coefficients that can be computed numerically. Further, time derivatives acting on the spin-coefficients can be substituted using the evolution equations in \cite{Beyer:2017}, given that these are written in terms of system variables and null rotation functions involved in the transformation to the Bondi frame. Performing a first order reduction then yields the equation for the Bondi time $u$ in the form
\begin{gather}
    \partial_t u = v, \nonumber \\
    \partial_t v = (K_1\eth + K_2\eth')u + 
    (C_1\eth^2 + C_2\eth\eth' + C_3\eth'\eth + C_4\eth'^2 + C_5\eth + C_6\eth' + C_7)v, \nonumber \\
    u(t_0) = 0,\qquad v(t_0)=1,
\end{gather}
where $C_i$ and $K_i$ are known numerically, the $\eth$-operators now represent those on the unit 2-sphere and $t=t_0$ represents the time of the first cut of $\mathscr{I}^+$ in the computational domain. This system is numerically evolved along $\mathscr{I}^+$ with a simple Euler step.

\subsection{The comparisons at a glance}\label{sec:ComparisonsAtAGlance}
In this subsection we outline some basic results before delving into more detail in the next subsection. We look at the hardest case to follow, namely the slowest oscillating modes given by $l=2$, which requires the longest physical proper time.

We specify the ingoing wave by setting the free data $q_0$ for the light-like ingoing characteristic variable on the outer boundary \cite{Beyer:2017}. First, we use an ingoing wave whose angular part is proportional to ${}_2Y_{20}$ specified at the outer boundary at $r=(5/2)m$
\begin{equation}
	q_0(t,\theta) =
	\begin{cases} 
	    40\sqrt{\frac{2\pi}{15}}\ii\sin^8(8{\pi t })\;{}_2Y_{20}(\theta)& t \leq\frac18 \nonumber \\
	    0 & t >\frac18
	\end{cases}.\label{eq:WaveProfile}
\end{equation}
The profile is chosen to create a smooth bump. Having the wave be proportional to ${}_2Y_{20}$ has the consequence of creating dynamics in all even $l$-modes, while all odd ones remain unperturbed. Of note is that the physical $\Psi_0$ generated on the boundary has a maximum amplitude of around $140$ in relativistic units, and thus, we do not merely nudge this black hole; we kick it.

\begin{figure}[H]
    \centering
    \begin{subfigure}[t]{0.3\textwidth}
        \centering
        \includegraphics[height=1.2in]{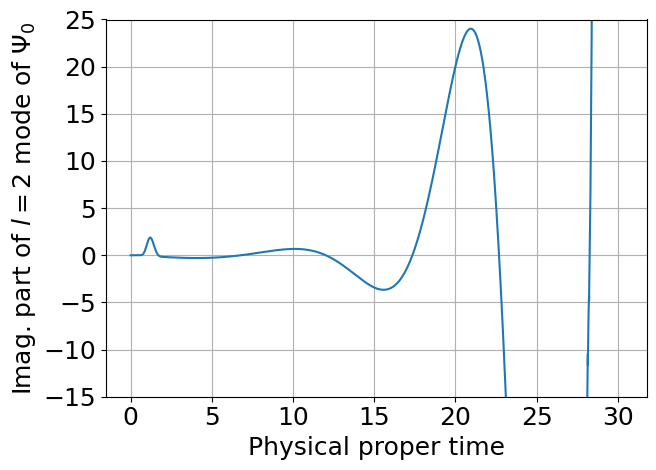}
        \caption{$\Psi_0$}
    \end{subfigure}%
    ~ 
    \begin{subfigure}[t]{0.3\textwidth}
        \centering
        \includegraphics[height=1.2in]{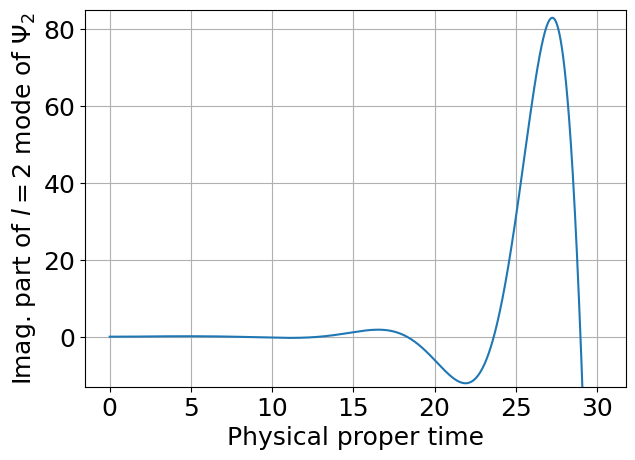}
        \caption{$\Psi_2$}
    \end{subfigure}
    ~
    \begin{subfigure}[t]{0.3\textwidth}
        \centering
        \includegraphics[height=1.2in]{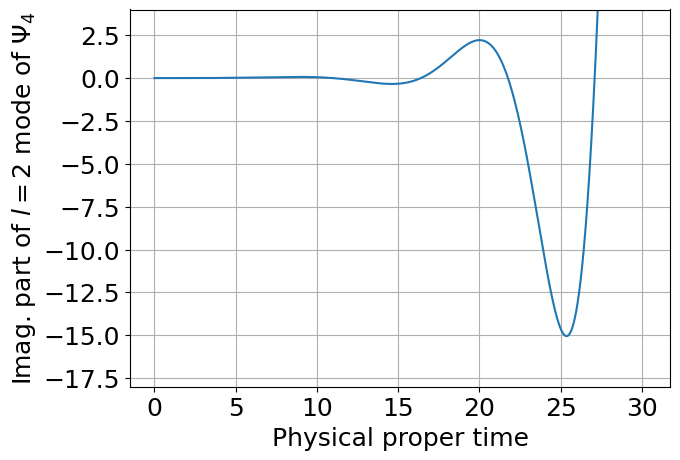}
        \caption{$\Psi_4$}
    \end{subfigure}
    \caption{The imaginary part of the $l=2$ spectral coefficient of $\Psi_0,\,\Psi_2,\,\Psi_4$ plotted with respect to proper physical time along the quasi-static world-line at $r\approx3.36$.}\label{fig:Diffpsi}
\end{figure}
\begin{figure}[H]
    \centering
    \begin{subfigure}[t]{0.5\textwidth}
        \centering
        \includegraphics[width=0.7\linewidth]{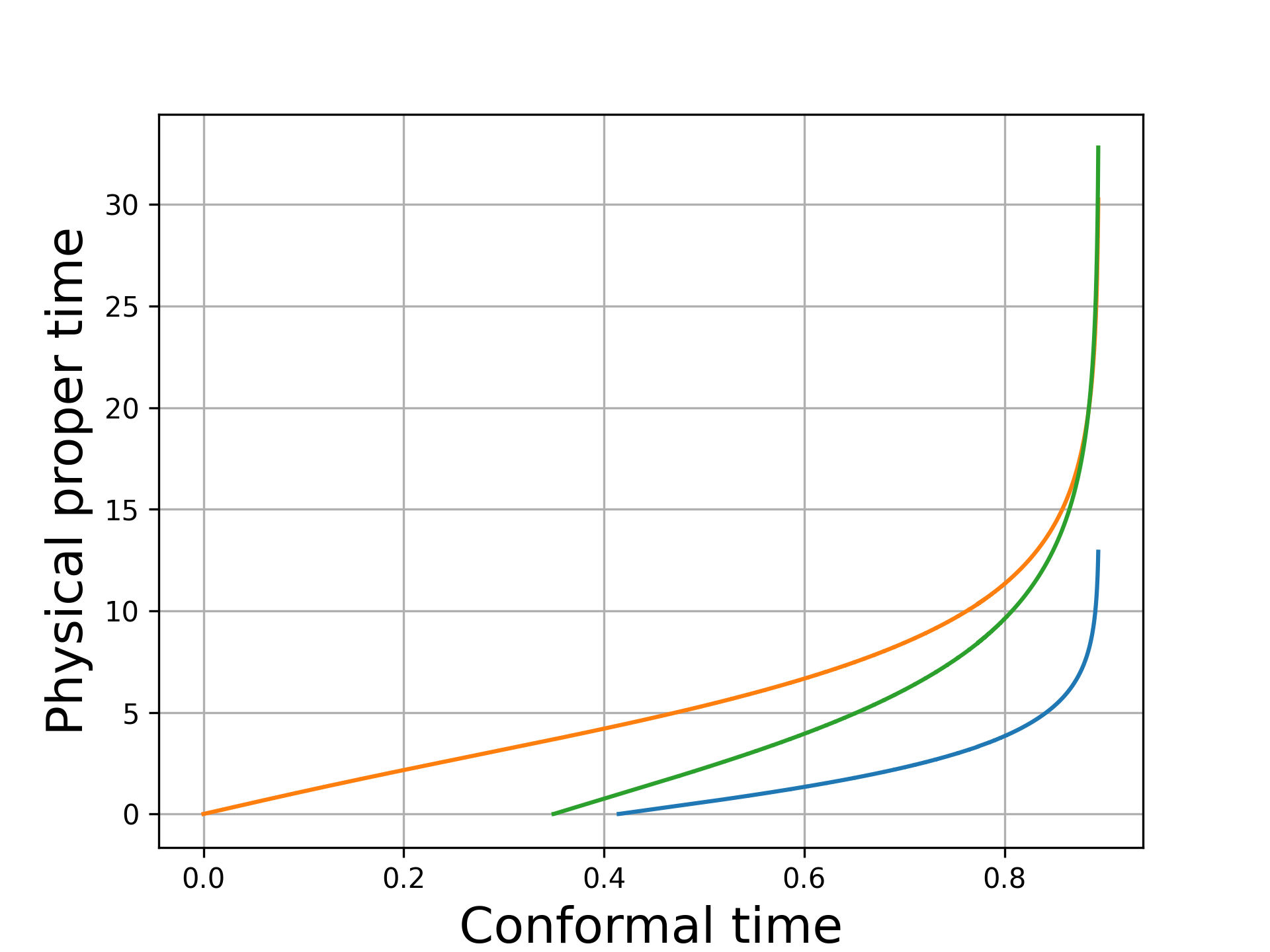}
        \caption{}
    \end{subfigure}%
    ~ 
    \begin{subfigure}[t]{0.5\textwidth}
        \centering
        \includegraphics[width=0.7\linewidth]{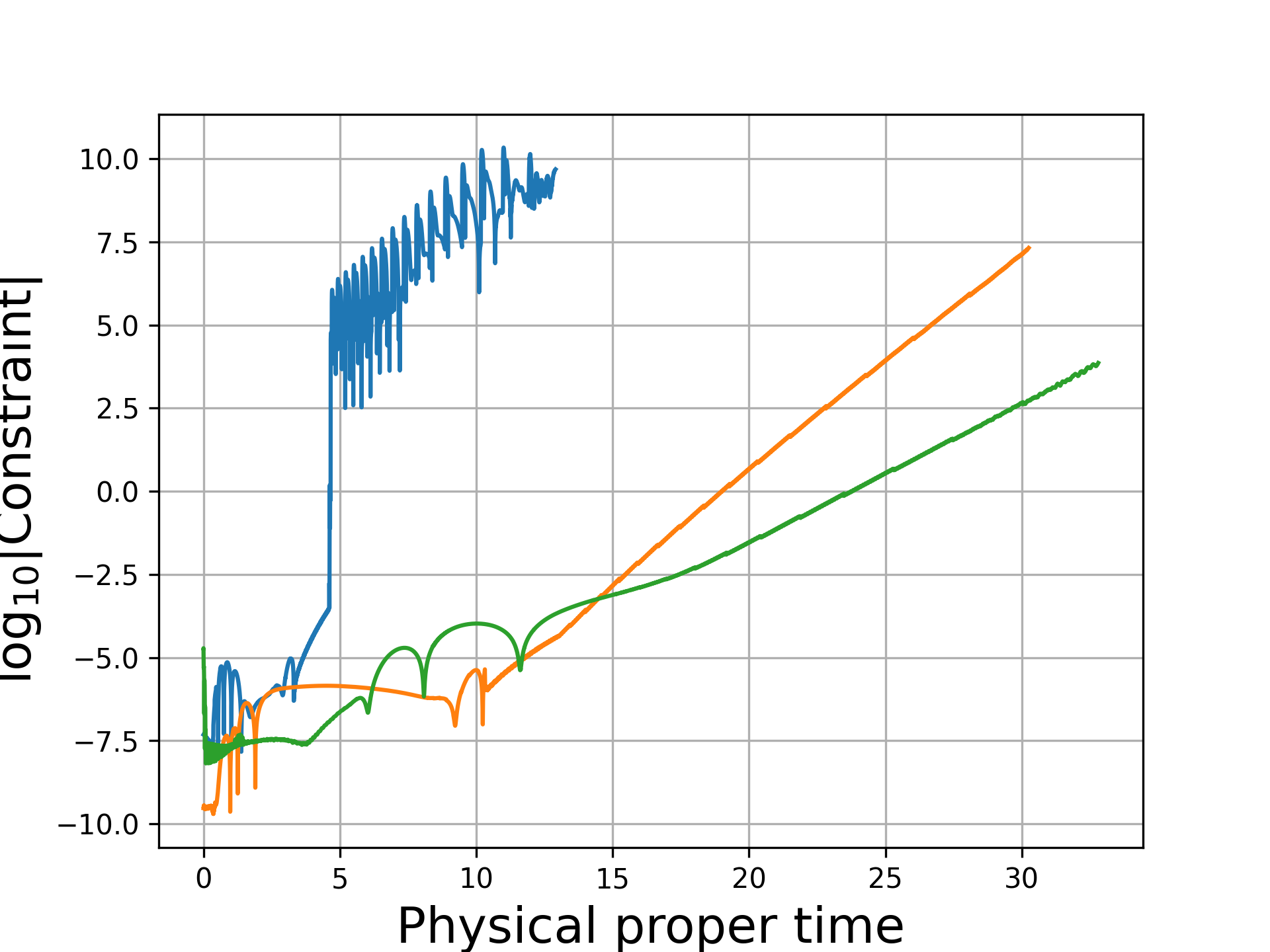}
        \caption{}
    \end{subfigure}
    \caption{(a) From bottom to top, the physical proper time along the quasi-static world-lines at $r\approx1.85,\,2.25$ and $10$. (b) $\log_{10}$ of the absolute value of constraint along the same curves as (a) for fixed angle $\theta=\pi/2$.}
    \label{fig:Constraint_and_PT}
\end{figure}

As a first look, Fig.~\ref{fig:Diffpsi} shows the induced oscillation in the imaginary part of the $l=2$ mode of the three non-vanishing complex components of the physical Weyl curvature $\Psi_{ABCD}$ along the quasi-static world-line at $r\approx3.31$ with angle $\theta=\pi/2$\footnote{This angle will be used throughout the paper as it corresponds to the maximum of the ingoing wave, thus giving the largest perturbation and easiest comparisons.}. Fig.~\ref{fig:Constraint_and_PT} shows the physical proper time along three such world-lines and a constraint coming from the Bianchi part of the constraint system to give an idea of accuracy. The violation of the constraints is worse the closer the curves are to the boundaries of the domain. This violation ultimately becomes significant inside the black hole when the regridding process to avoid the singularity puts the boundary ever closer to the world-line.

\begin{figure}[H]
    \centering
    \begin{subfigure}[t]{0.3\textwidth}
        \centering
        \includegraphics[height=1.2in]{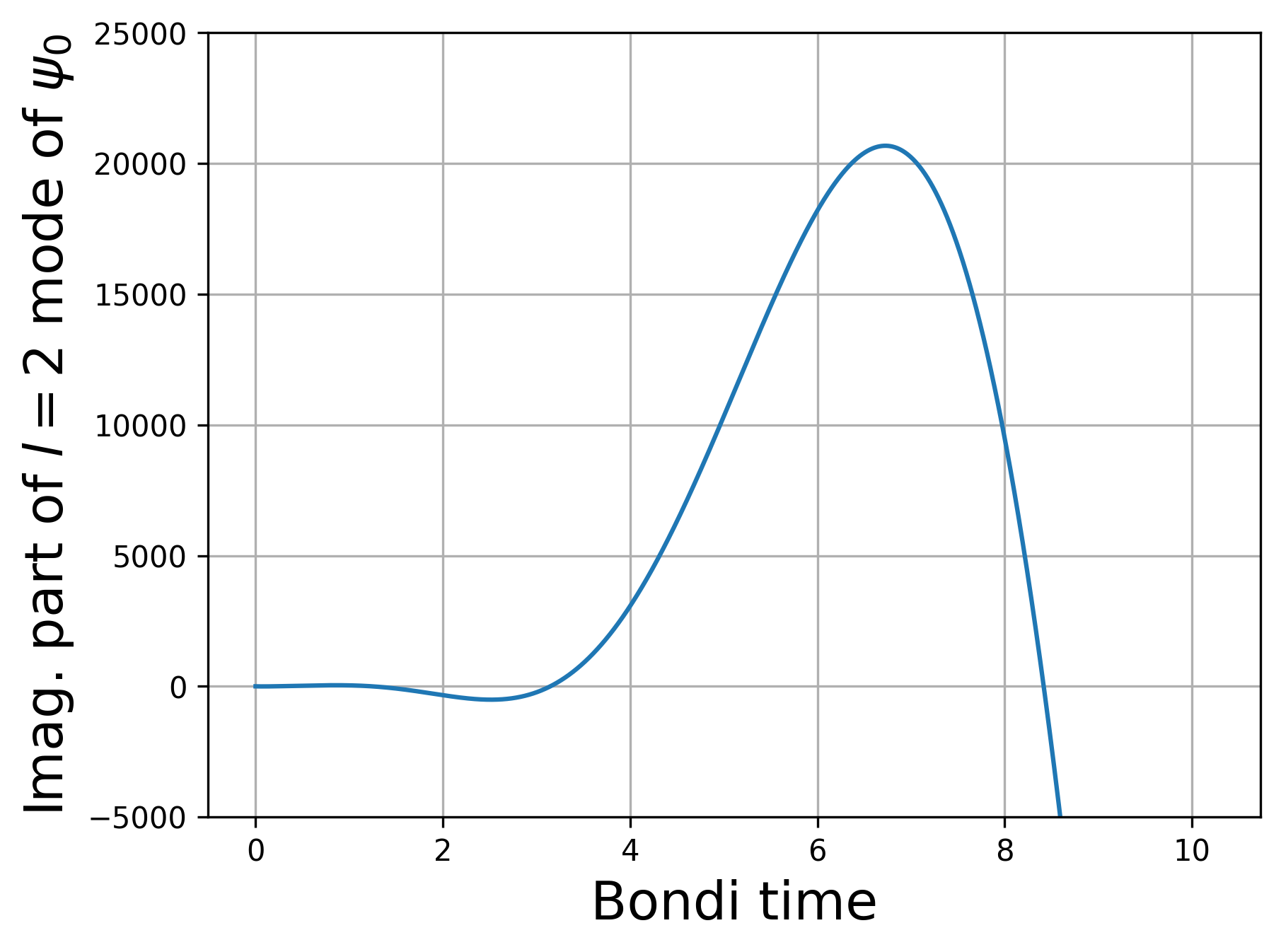}
        \caption{$\psi_0$}
    \end{subfigure}%
    ~ 
    \begin{subfigure}[t]{0.3\textwidth}
        \centering
        \includegraphics[height=1.2in]{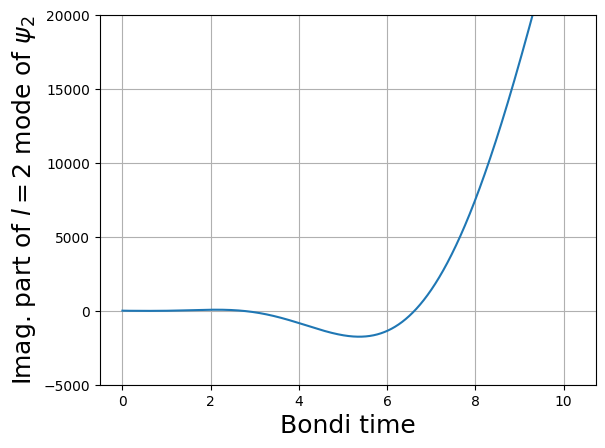}
        \caption{$\psi_2$}
    \end{subfigure}
    ~
    \begin{subfigure}[t]{0.3\textwidth}
        \centering
        \includegraphics[height=1.2in]{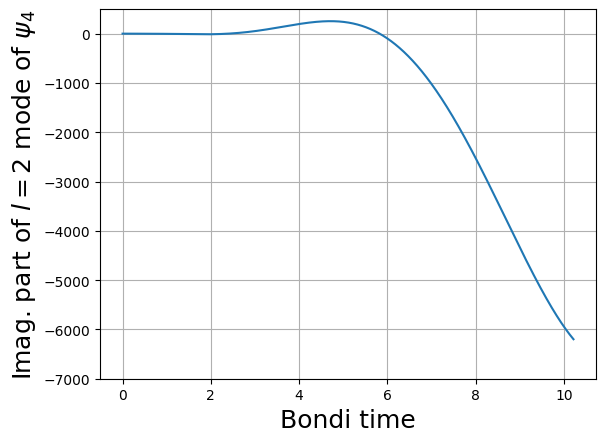}
        \caption{$\psi_4$}
    \end{subfigure}
    \caption{The imaginary part of the $l=2$ spectral coefficient of some components of $\psi_{ABCD} = \Theta^{-1}\Psi_{ABCD}$ plotted with respect to Bondi time along $\mathscr{I}^+$.}\label{fig:Diffpsiscri}
\end{figure}

\begin{figure}[H]
    \centering
    \begin{subfigure}[t]{0.5\textwidth}
        \centering
        \includegraphics[width=0.7\linewidth]{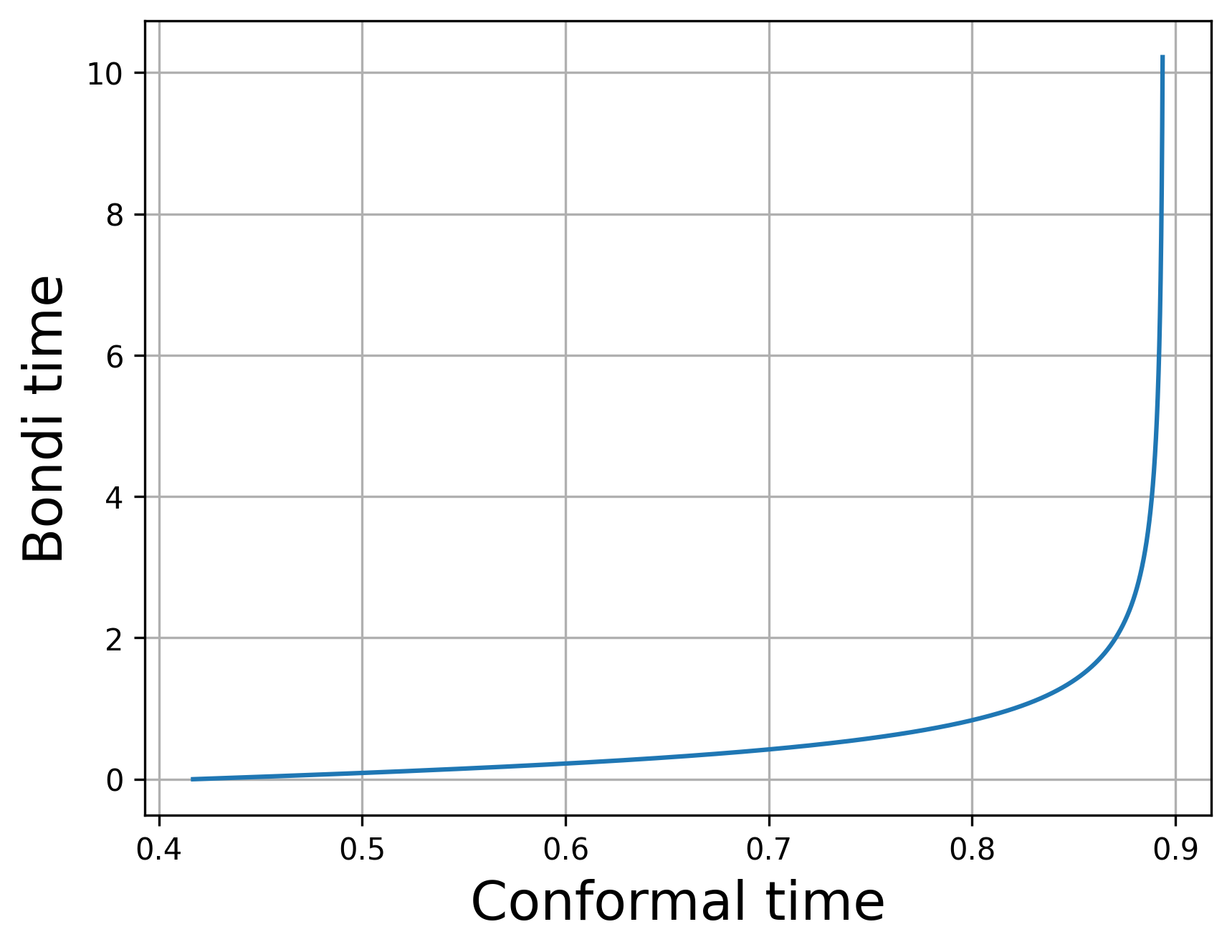}
        \caption{}
    \end{subfigure}%
    ~ 
    \begin{subfigure}[t]{0.5\textwidth}
        \centering
        \includegraphics[width=0.7\linewidth]{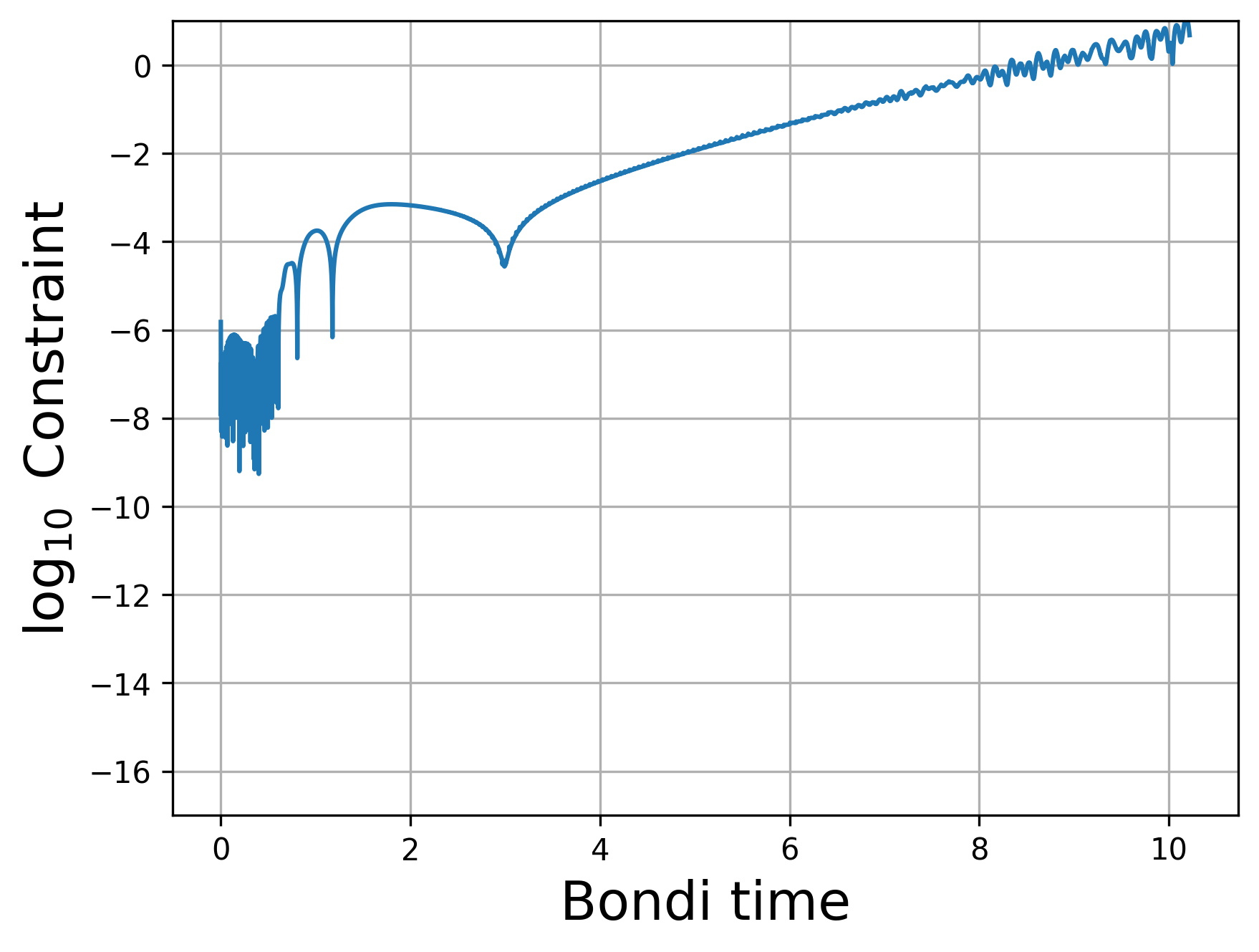}
        \caption{}
    \end{subfigure}
    \caption{(a) The Bondi time along a null generator of $\mathscr{I}^+$. (b) $\log_{10}$ of the absolute value of a Bianchi constraint along the curve in (a) for fixed angle $\theta=\pi/2$.}
    \label{fig:Constraint_and_BT}
\end{figure}

Figs.~\ref{fig:Diffpsiscri} and~\ref{fig:Constraint_and_BT} present analogous plots but now on $\mathscr{I}^+$ and components of the rescaled Weyl spinor $\psi_{ABCD}$ with respect to  Bondi time.


Figs.~\ref{fig:contourplot}--\ref{fig:Constraint_and_BT} give a nice overview of how the conformal field equations  with the conformal Gauß gauge perform when attempting to follow physical phenomena in physical proper time. It is clear there are issues when attempting this out-of-the-box. The compactification is too fast for any numerical regime to resolve anything after only 10 to 20 units of physical proper time. The infinite physical proper distance between the black hole's horizon and $\mathscr{I}^+$ is squashed into a tiny amount of the conformal space-time's radial coordinate. The infinite physical proper time between a value of 10-20 or so and $i^+$ is squashed into a tiny amount of the conformal space-time's temporal coordinate. Such a tight compactification, containing so much physical dynamics, is practically impossible for a numerical framework to resolve, no matter how high the resolution.

It is also apparent that there are not as many oscillations along $\mathscr{I}^+$ as there are along the quasi-static world-lines inside the space-time. This property can be very roughly deduced by considering the contour plot Fig.~\ref{fig:contourplot}, which should be noted is not a conformal diagram; null curves are not 45 degree lines. One can see roughly the trajectory of ingoing null curves by noting the path of the ingoing wave. Similarly, one can see a very rough approximation to the outgoing null curves by noting the path of our approximation to the apparent horizon. Picking some point in the physical space-time and then (roughly) tracing a null curve to $\mathscr{I}^+$ will inevitably lead to a larger conformal temporal and radial value. Thus, we must go further in conformal time along $\mathscr{I}^+$ to reproduce the same curvature oscillations seen along quasi-static world-lines at earlier conformal times.

Of importance here is that the code is barely able to evolve far enough in physical proper time to resolve the slowest curvature oscillations before the compactification becomes impossible to handle directly. This is exemplified in the computational domain shrinking to an impractically small size with respect to the conformal coordinates as $i^+$ is approached, resulting in exceedingly bad violations of the constraints. Further, as $i^+$ is a singular point, the system variables $\psi_i$ diverge there as well. Thus, there are many contributing factors to the simulation crashing. Importantly, this happens after only a short amount of physical proper time has passed.

Before giving an overview of procedures aimed at solving these problems (see Sec.~\ref{sec:ProblemsandResolutions}), we provide, in the next subsection, a more detailed analysis of comparisons to linear theory in the present framework.

\subsection{A detailed comparison}\label{sec:DetailedComparisons}

The first property one should check is how close the frequencies of the curvature oscillations are to the frequencies given by linear theory. We expect that the non-linear effects of a perturbed black hole die down very quickly, and thus the change in frequency over time should approach the linear regime quickly. In linear theory, the perturbed $l$-modes are completely independent of one another so that the SWSH modes can be excited individually. Here, due to the  non-linearity of the evolution system, perturbing the space-time with an ingoing gravitational pulse proportional to a single mode, such as the $l=2$ SWSH for example (as done in the previous section) results in the excitation of all other even $l$-modes, which is rather convenient. The excitation of these other modes is a non-linear effect.

\begin{table}[H]
  \centering
        \begin{tabular}{ |c|c|c|c|c| } 
        \hline
            {} & Lin. Freq. & $E_1$ & $E_2$ & $E_3$ \\ 
            \hline
            $\Im(\Psi_0)_{l=2}$ & 0.58720 & 11.6\% & 2.11\% & 0.322\% \\ 
            \hline
            $\Im(\Psi_0)_{l=4}$ & 1.27157 & 11.8\% & 19.3\% & 17.6\% \\ 
            \hline
            $\Im(\Psi_4)_{l=2}$ & 0.58720 & 7.25\% & 1.28\% & 0.278\% \\ 
            \hline
            $\Im(\Psi_4)_{l=4}$ & 1.27157 & 61.6\% & 14.1\% & 13.1\% \\ 
            \hline
        \end{tabular}
        \caption{\label{tab:l2pert}A comparison along the quasi-static world-line at $r\approx3.31$ of the non-linear curvature oscillations to linear theory. The frequency at each half period has relative error '$E_i$' to the frequency, 'Lin. Freq.', from linear theory for an ingoing gravitational pulse proportional to ${}_2Y_{20}$.}
    \end{table}

Tab.~\ref{tab:l2pert} compares the frequency of the imaginary $l=2$ SWSH modes of $\Psi_0$ and $\Psi_4$ calculated at each half period to the frequency given by linear theory. It is clear that the modes corresponding to the original gravitational perturbation resemble lineary theory the closest.

\begin{table}[H]
  \centering
%  \small
  \begin{tabular}{ |c|c|c|c|c|c|c| } 
    \hline
    {} & Lin. Freq. & $E_1$ & $E_2$ & $E_3$ & $E_4$ & $E_5$ \\ 
    \hline
    $\Im(\Psi_0)_{l=2}$ & 0.58720 & 184.5\% & 53.2\% & 76.4\% & 117.5\% & 73.9\%\\ 
    \hline
    $\Im(\Psi_0)_{l=4}$ & 1.27157 & 2.20\% & 1.10\% & 0.344\% & 0.616\% & 0.156\% \\ 
    \hline
    $\Im(\Psi_4)_{l=2}$ & 0.58720 & 197.4\% & 50.8\% & 90.7\% & 85.7\% & 88.5\% \\ 
    \hline
    $\Im(\Psi_4)_{l=4}$ & 1.27157 & 1.22\% & 0.306\% & 0.412\% & 0.808\% & 0.052\%  \\ 
    \hline
  \end{tabular}
  \caption{\label{tab:l4pert}A comparison along the quasi-static world-line at $r\approx3.31$ of the non-linear curvature oscillations to linear theory. The frequency at each half period has relative error '$E_i$' to the frequency, 'Lin. Freq.', from linear theory for an ingoing gravitational pulse proportional to ${}_2Y_{40}$.}
\end{table}
Tab.~\ref{tab:l4pert} compares frequencies as in Tab.~\ref{tab:l2pert} but now with an ingoing gravitational pulse proportional to ${}_2Y_{40}$. It is clear that the $l=4$ modes now approach the linear regime quicker than the $l=2$ modes. This indicates the non-linearities that appear in modes different from the initial perturbation, are slower to enter the linear regime. It is quite remarkable that within a few periods the directly excited modes are very well described by linear theory.

\subsection{Results outside standard linear theory}
\begin{figure}[htb]
    \centering
    \begin{subfigure}[t]{0.33\textwidth}
        \centering
        \includegraphics[width=0.8\linewidth]{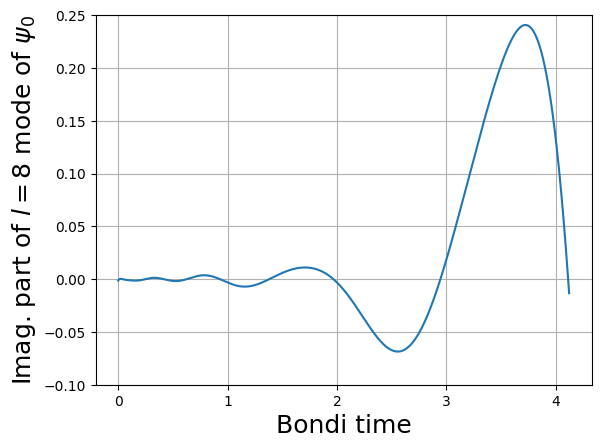}
        \caption{$\psi_0$}
    \end{subfigure}\quad
    \begin{subfigure}[t]{0.33\textwidth}
        \centering
        \includegraphics[width=0.8\linewidth]{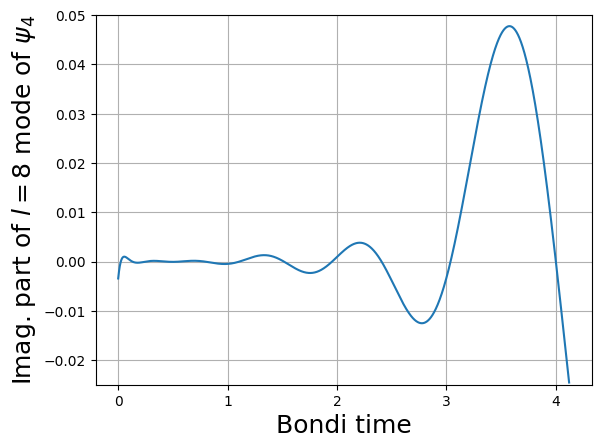}
        \caption{$\psi_4$}
    \end{subfigure}\quad
    \begin{subfigure}[t]{0.33\textwidth}
        \centering
        \includegraphics[width=0.8\linewidth]{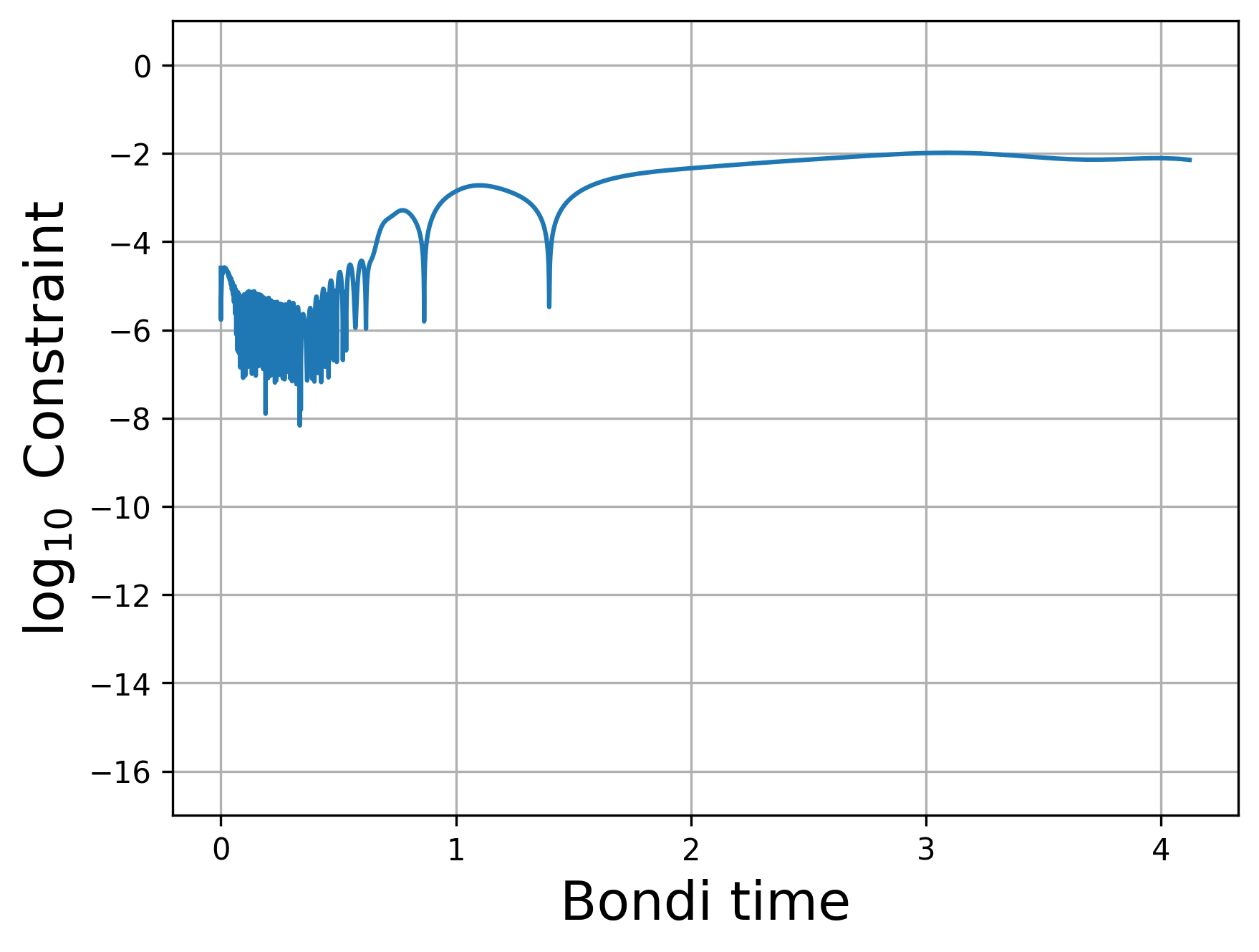}
        \caption{$\log_{10}|\text{Constraint}|$}
    \end{subfigure}
    \caption{The imaginary parts of the $l=8$ modes of (a) $\psi_0$ and (b) $\psi_4$ with respect to Bondi time along $\mathscr{I}^+$. (c) Shows $\log_{10}$ of a constraint coming from the Bianchi subsystem.}
    \label{fig:l8Pertbs}
\end{figure}

Other than comparing our curvature oscillations to linear theory, we can also see the behaviour of the Weyl scalars on curves inside the black hole and along $\mathscr{I}^+$. The frequency along $\mathscr{I}^+$ with respect to  Bondi time should be identical to that along the quasi-static world-line with respect to  Schwarzschild time~\cite{Zenginoglu:2009}. Due to the oscillations being delayed on $\mathscr{I}^+$ as compared to quasi-static world-lines at the same conformal time, we do not resolve as many oscillations. However, when shooting in an $l=4$ gravitational pulse, the higher order modes have considerable amplitude, see Fig.~\ref{fig:l8Pertbs}, by comparison of their size to the level of satisfication of the constraint equations; this gives evidence that they are accurately resolved. Further, the higher order modes have a higher frequency, giving us more rings. As Tab.~\ref{tab:l8pertscri} shows,  the non-linear effect here is much larger than with modes closer to the excited mode, however there is a clear trend toward the linear regime.

\begin{table}[H]
  \centering
        \small
        \begin{tabular}{ |c|c|c|c|c|c|c|c|c|c| } 
        \hline
            {} & Lin. Freq. & $E_1$ & $E_2$ & $E_3$ & $E_4$ & $E_5$ & $E_6$ & $E_7$ & $E_8$ \\ 
            \hline
            $\Im(\Psi_0)_{l=8}$ & 1.606202 & 1335\% & 1024\% & 910.5\% & 652.3\% & 426.4\% & 255.6\% & 130.5\% & 67.64\% \\ 
            \hline
            $\Im(\Psi_4)_{l=8}$ & 1.606202 & 1053\% & 908.1\% & 599.2\% & 439.0\% & 366.2\% & 330.7\% & 244.5\% & 144.3\% \\
            \hline
        \end{tabular}
        \caption{\label{tab:l8pertscri}Along the generator of $\mathscr{I}^+$ starting at angle $\theta=\pi/2$ on the first cut of $\mathscr{I}^+$ we compare the non-linear curvature oscillations to linear theory. $E_i$ represent the relative error of the $l=8$ mode's frequency at each half period to the frequency from linear theory for an ingoing gravitational pulse proportional to ${}_2Y_{40}$.}
    \end{table}

    We can also look at curvature oscillations along curves inside the black hole. The standard approach in linear perturbation theory has a boundary at the horizon and thus excludes this region completely. However, as the contour plot Fig.~\ref{fig:contourplot} shows, curvature oscillations occur inside the black hole as well.

\begin{figure}[htb]
    \centering
    \begin{subfigure}[t]{0.5\textwidth}
        \centering
        \includegraphics[width=0.7\linewidth]{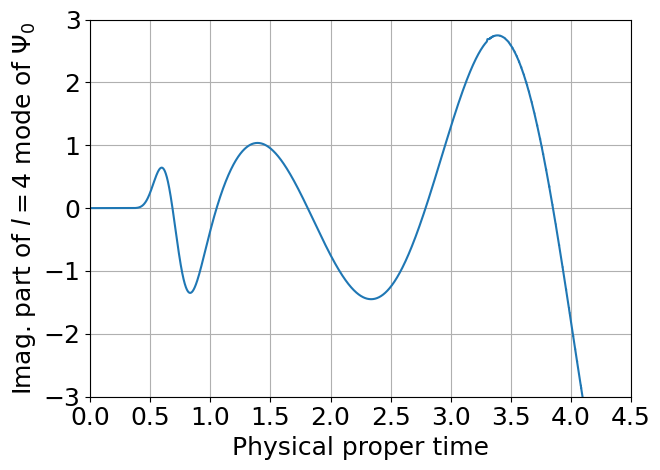}
        \caption{$\psi_0$}
    \end{subfigure}%
    ~ 
    \begin{subfigure}[t]{0.5\textwidth}
        \centering
        \includegraphics[width=0.7\linewidth]{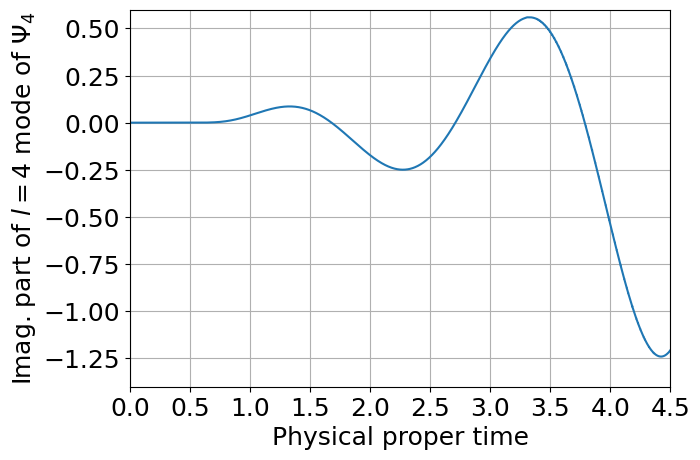}
        \caption{$\psi_4$}
    \end{subfigure} \\
    ~
    \begin{subfigure}[t]{0.5\textwidth}
        \centering
        \includegraphics[width=0.7\linewidth]{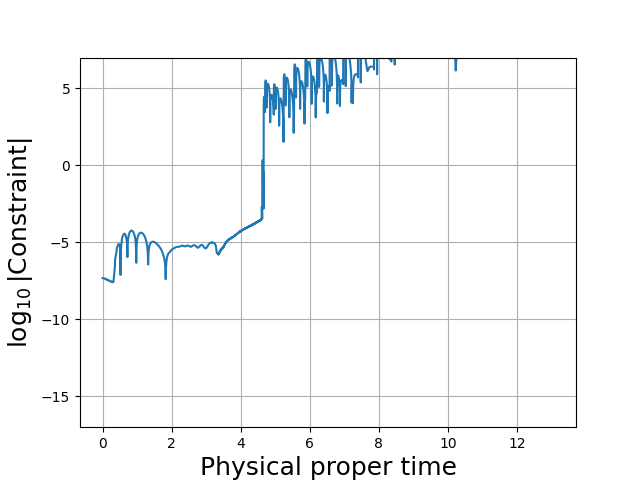}
        \caption{$\log_{10}|\text{Constraint}|$}
    \end{subfigure}
    \caption{The imaginary parts of the $l=4$ modes of (a) $\psi_0$ and (b) $\psi_4$ with respect to physical proper time along the quasi-static world-line at $r\approx1.85$. (c) Shows $\log_{10}$ of a constraint coming from the Bianchi subsystem.}
    \label{fig:l4PertinsideBH}
\end{figure}
One can see from Fig.~\ref{fig:l4PertinsideBH} that when the physical proper time is around $5$, the constraints become badly violated. This is due to the inner boundary coming too close to the quasi-static world-line, and the boundary conditions are inconsistent with the back-reacting ingoing gravitational wave. Tab.~\ref{tab:l4pertinsideBH} compares the frequency of the imaginary part of the $l=4$ modes of $\psi_0$ and $\psi_4$ before this time. It is clear that they quickly approach those given by linear theory.
\begin{table}[htb]
  \centering
  \begin{tabular}{ |c|c|c|c|c| } 
    \hline
    {} & Lin. Freq. & $E_1$ & $E_2$ & $E_3$ \\ 
    \hline
    $\Im(\Psi_0)_{l=4}$ & 2.841916 & 97.15\% & 16.84\% & 7.15\% \\ 
    \hline
    $\Im(\Psi_4)_{l=4}$ & 2.841916 & 17.21\% & 6.78\% & 0.75\% \\
    \hline
  \end{tabular}
  \caption{\label{tab:l4pertinsideBH}Along the quasi-static world-line at $r\approx1.85$ we compare the non-linear curvature oscillations to linear theory. The frequency at each half period has relative error '$E_i$' to the frequency, 'Lin. Freq.', from linear theory for an ingoing gravitational pulse proportional to ${}_2Y_{40}$.}
\end{table}

The results of this section show that we cannot go very far in physical proper time with the GCFE as currently implemented. We can, however, deduce that what we are seeing agrees with previous results in that non-linear curvature oscillations quickly conform to the linear regime. The modes that approach the linear regime the quickest are those matching the ingoing wave. Those which do not, exist due to  non-linear effects, and take longer to approach the linear regime.

We have also computed modes with respect to Bondi time along generators of $\mathscr{I}^+$ as well as for what would be equivalent to quasi-static observers inside the black hole. Both of these regimes are excluded by standard QNM calculations, but we have seen that the approximate frequencies also approach the values for the linear theory in these cases.

\section{Problems with the compactification}\label{sec:ProblemsandResolutions}

Here we briefly lay out the different issues faced in imposing a constant-in-physical-proper-time timstep with the conformal Gauß gauge.

In~\cite{Frauendiener:2021a}, it was touched upon that the compactification procedure while in the conformal Gauß  gauge acts extremely fast with respect to proper physical time $\tau$. This has been exemplified in the previous sections by attempting to resolve as many curvature oscillations as possible. The compactification problem is most easily exemplified by choosing a time-like conformal geodesic that the gauge is built upon. The spatial coordinates, by definition of the gauge, are constant along this curve. Therefore, the proper physical time $\tau$ along this curve is related to the conformal time by $\text{d}\tau^2 = \Theta^{-2}\text{d}t^2$. Since we know that the conformal factor satisfies $\partial_t^3\Theta = 0$, we can express it as a quadratic function in $t$
\begin{equation}
    \Theta(t) = c + 2 b t + a t^2,
\end{equation}
for some $a,\,b,\,c$ that are functions of the spatial coordinates. This is one of the marvels of the conformal Gauß gauge: the conformal factor is known \emph{a priori}. Integration for $\tau$ immediately yields (up to an integration constant)
\begin{equation}\label{eq:PT}
    \tau(t) =
    \begin{cases} 
      -\frac{1}{f}\text{arctanh}\Big{(}\frac{at+b}{f}\Big{)} & f^2 > 0, \\
     - \frac{1}{at+b} & f^2 = 0, \\
        \frac{1}{f}\text{arctan}\Big{(}\frac{at+b}{f}\Big{)} & f^2 < 0
 \end{cases}\qquad f^2 := b^2 - ac.
\end{equation}
Thus, there are two very different possible relationships between $\tau$ and $t$, depending on the sign of $f^2$. For asymptotically flat space-times, two real roots of the conformal factor are needed in order to compactify future and past null-infinity, so necessarily $f^2 > 0$. In this case, the conformal time when this occurs is $t_{\pm}=\pm(f-b)/a$. For the gauge choice and space-time considered here, a time-like conformal geodesic that starts at $t=0$ from the special radius $r=(3 + \sqrt{5})m/4 \approx 1.309$ will pass through future and past time-like infinity $i^{\pm}$ at $t_{\pm} = \pm 2/\sqrt{5} \approx \pm 0.894$~\cite{Friedrich:2003} (compare with Fig.~\ref{fig:contourplot}).

\begin{figure}[htb]
    \centering
    \begin{subfigure}[t]{0.47\textwidth}
        \centering
        \includegraphics[width=0.7\linewidth]{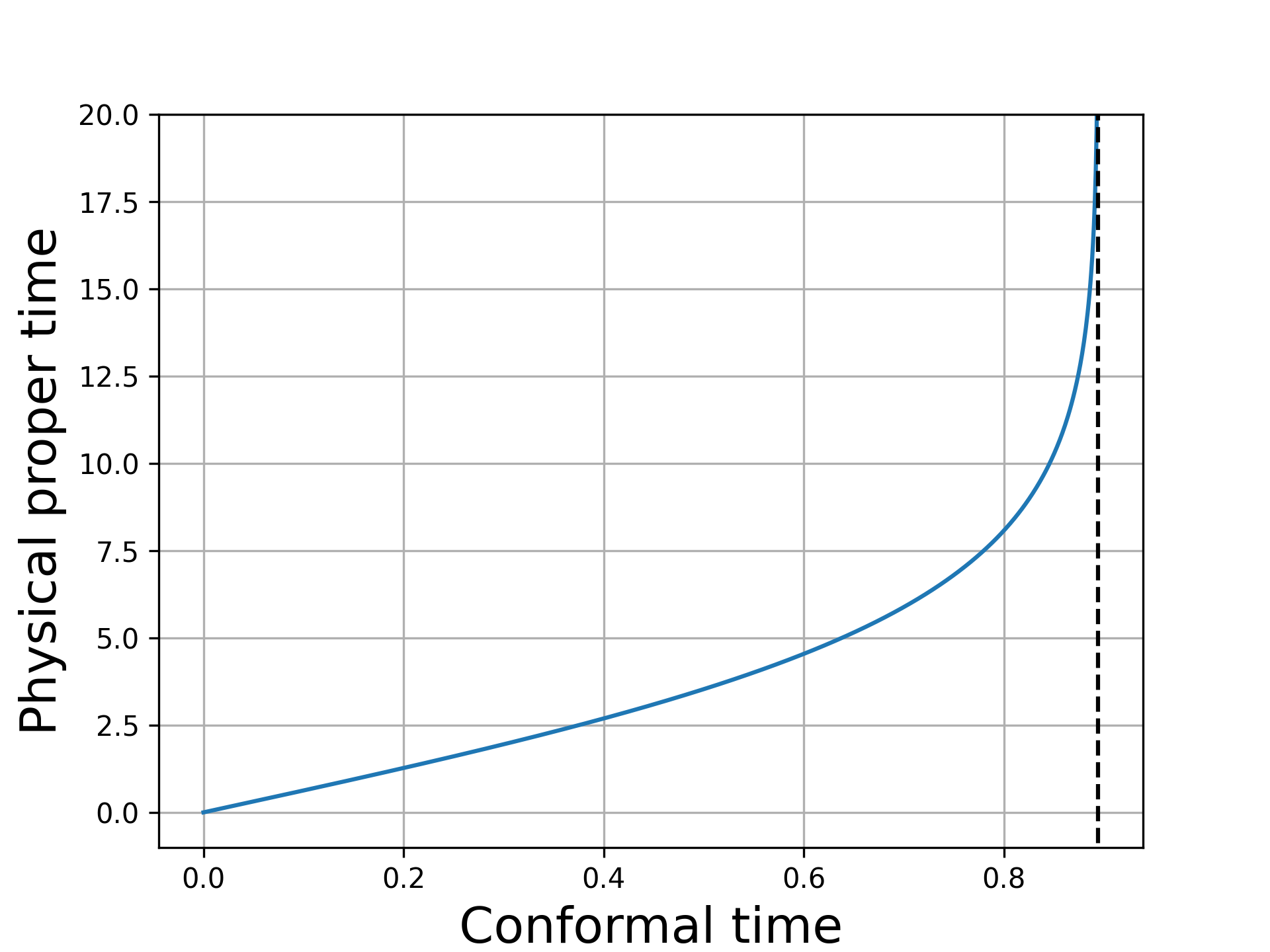}
        \caption{}
    \end{subfigure}
    ~ 
    \begin{subfigure}[t]{0.47\textwidth}
        \centering
        \includegraphics[width=0.7\linewidth]{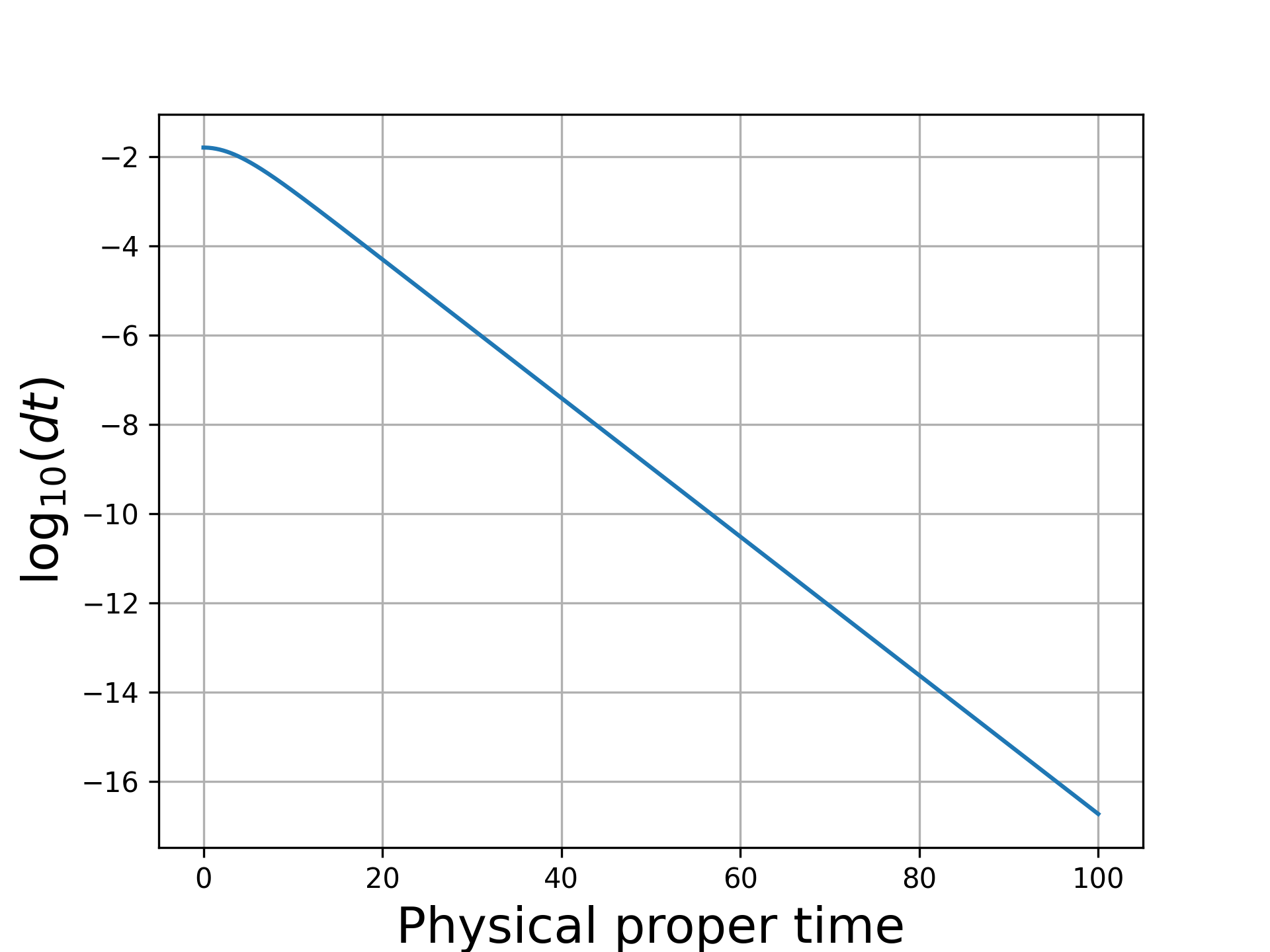}
        \caption{}
    \end{subfigure}
    \caption{Looking along the time-like conformal geodesic that intersects $i^{\pm}$ at $t\approx0.894$ we have (a) the relationship between physical proper time $\tau$ and conformal time $t$ and (b) the $\log_{10}$ of the conformal timestep required to attain a constant physical proper timestep of 0.1 with respect to proper physical time.}
    \label{fig:PT}
\end{figure}

The issue that confronts us is how far we can step in physical proper time that is numerically tractable. Fig.~\ref{fig:PT} shows that along the time-like conformal geodesic that intersects $i^+$ we have an extremely large slope after evolving up to 20 units of physical proper time. Thus, imposing a timestep of 0.1 in physical proper time --- a practical choice to resolve the curvature oscillations (although not with very good resolution)--- the timestep in conformal (code) time will fall below the machine precision for 64-bit floats when $\tau$ is around 100. This is completely in line with what we find in Sec.~\ref{sec:DetailedComparisons} in attempting to follow the curvature oscillations in physical proper time.

A similar issue concerns the radial compactification at late times. The conformal ``size'' of the region  between the black hole's horizon and null-infinity becomes extremely small while the number of grid points is maintained due to the regridding procedure we apply. Again, this results in the ultimate underflow of the interval between the points.

%It is worthwhile commenting briefly on what happens when $f < 0$, which is the case for a standard compactification of the Anti-de Sitter space-time as a submanifold of the Einstein cylinder. This space-time has a time-like conformal boundary and necessarily $f < 0$. In this case, the arctan behaviour of the conformal factor is the opposite of the previous situation. There is no problem with the conformal time parameter running to infinity, however the physical proper time parameter will asymptote to a finite value as this happens. The problem in this case is that one does not go far enough, and something must be done to extend the amount of physical time traversed.

It is clear, that there are issues of a principle nature when an infinite system is represented on a finite grid. These issues are particularly manifest in our present study of quasi-normal modes which requires a stretch of physical time for transient non-linear effects to die down before the linear regime is entered. Previous work on the numerics of the conformal field equations (see~\cite{Hubner:2001a,Frauendiener:1998a}) suggest that there are gauge choices for which the spatial compactification can be prevented ('scri'-freezing shift condition) and singularity avoiding gauges for which the approach to time-like infinity can be slowed down. Thus, the issues we have discussed above are due to our choice of the conformal Gauß gauge.

One might then suggest to simply choose a different gauge. However, this is easier said than done since the conformal Gauß gauge simplifies the conformal field equations drastically. As we have seen in Sec.~\ref{sec:system} most fields are simply carried along the time-like conformal geodesics, the only exception being the rescaled Weyl curvature which propagates along the light-cone, a time-like cone and the time lines. In a different gauge this will change with the consequence that the analysis of the boundary conditions at the time-like boundaries will become much more complicated. Since this is a highly non-trivial and complicated procedure we do not consider this option.

A potential resolution, while remaining in the conformal Gauß gauge, is to reparametrize the time-like conformal geodesics before the compactification becomes too fast, and then rebuild the gauge. The reparametrization can be chosen to change the relationship between the conformal time and physical proper time in such a way that the extreme compactification is postponed. In principle, this procedure can be applied any number of times, allowing one to proceed as far as one wants in proper physical time while simultaneously maintaining a manageable transformation to conformal time. Performing a temporal reparametrisation while maintaing the conformal Gauß gauge results in not only a new temporal coordinate, but a new orthonormal frame and conformal factor as well.

%Thus, when transforming to a new choice of conformal Gauß gauge, one could either choose a new temporal coordinate or a new conformal factor; one can arrive at the same new gauge through either path.

% The conformal Gau\ss\;gauge could be slightly generalized by not requiring that the temporal coordinate is a preferred parameter. Instead, one could use the more general $u^a(x^i)\nabla_at=N(x^i)$, where $u^a$ is the tangent vector to the conformal geodesics, $t$ the temporal parameter, $N(x^i)$ a lapse and $x^i$ are the spatial coordinates. The choice $N=1$ would recover the conformal Gau\ss\;gauge. This would add the ability to specify the speed of $t$ along the conformal geodesics and may be of use, so long as this does not interfere with the characteristics of the system.

\section{Summary and future work}\label{sec:summary}

We have investigated the viability of the generalized conformal field equations, written in the conformal Gauß gauge, for evolving space-times while maintaining a constant physical proper timestep. To study this, we shot an axi-symmetric gravitational pulse into a Schwarzschild black hole and followed the curvature oscillations that were subsequently produced. These are well documented in the linear regime as quasinormal modes and ring at a constant frequency along curves of constant Schwarzschild radius with respect to Schwarzschild (or proper physical) time. %They were then an excellent test.

We put forward the notion of quasi-static observers, which are points at a fixed angle on a temporal family of 2-surfaces with constant area, mimicking curves of constant Schwarzschild radius in our non-linear setting. They coincide with those when there is no gravitational perturbation. We also considered idealised observers at infinity with respect to Bondi time which we calculated by solving a second order differential equation along the generators of $\scri$. Thus, three distinct curves were eventually discussed, the first two being quasi-static observers inside and outside the black hole, with the third being an observer at infinity. Only the curve outside the black hole has a correspondence in linear theory, since the other two are inaccessible by construction in the linear setting.

Along these curves we looked at different spin-weighted spherical harmonic modes of the Weyl scalars $\Psi_i$ for curves in the physical space-time, and the conformal Weyl scalars $\psi_i$ for the curves along null-infinity, where the gravitational perturbation was proportional to ${}_2Y_{20}$ or ${}_2Y_{40}$. Although our perturbation is proportional to just one mode, because we are in the fully non-linear regime, all other even $l$-modes were excited. We found that the directly excited mode of the Weyl scalars was the quickest to approach the linear regime. This happened very quickly, within a few periods, after which the relative error to the linear theory's frequency was less than $1\%$. The non-linear parts of the perturbation, namely the modes differing from the ingoing perturbation's, were approaching the linear regime as well, but at a slower rate.

With regards to the viability of the conformal Gauß gauge for this study, we found that after a short amount of physical proper time, of the order of 20 units or so, the compactification was so fast, that the remaining physical space-time was compressed into a very small temporal and radial region with respect to the conformal coordinates. This has, in particular, the consequence of requiring a very small conformal timestep to resolve the curvature oscillations. It was found that this became too much for the numerics and the simulation stopped.

There are avenues toward reigning in the compactification through choosing a different conformal factor or resetting the temporal coordinate. These two procedures are in fact equivalent. The outcome would be a complete gauge transformation involving coordinates, frame and conformal factor that would induce a transformation of all the system variables.

As a final point we should point out that we have restricted our consideration to the case where the frame was rigidly tied to the conformal geodesics. This does not have to be the case. One could conceive of the possibility to choose the frame in a different way along the conformal geodesics. This would introduce a freedom which is akin to the lapse and shift freedom in the choice of coordinates. This might then open up the possibility for implementing 'scri'-freezing shift conditions as well as a singularity avoiding time coordinate as was explored in~\cite{Frauendiener:1998a}. However, it is very likely that this will also destroy the remarkable simplicity of the system of equations that we are solving. Investigations into these gauge issues will appear in future work.

As always, there is a trade-off between various objectives and the best solution for a particular objective must be found. We have shown that the GCFE with the conformal Gauß gauge are excellent for computing global properties of space-times such as the Bondi-Sachs massloss~\cite{Frauendiener:2021a} and the Newman-Penrose constants that we will discuss in an upcoming paper.

\ack
This work was supported by the Marsden Fund Council from Government funding, managed by Royal Society Te Apārangi.

The authors wish to acknowledge the use of New Zealand eScience Infrastructure (NeSI) high performance computing facilities, consulting support and/or training services as part of this research. New Zealand's national facilities are provided by NeSI and funded jointly by NeSI's collaborator institutions and through the Ministry of Business, Innovation \& Employment's Research Infrastructure programme. URL https://www.nesi.org.nz.

%\appendix

% \section*{References}
\printbibliography
% \bibliographystyle{elsarticle-num}
% \bibliographystyle{cqg}
% \bibliography{QNMbib.bib}
\end{document}